\documentstyle[12pt]{article}

\def\be{\begin{equation}}
\def\ee{\end{equation}}
\def\bea{\begin{eqnarray}}
\def\eea{\end{eqnarray}}
\def\ba{\begin{array}}
\def\ea{\end{array}}
\def\ben{\begin{enumerate}}
\def\een{\end{enumerate}}

\def\nnu{\nonumber}
\def\ll{\label}
 
\begin{document}
\newcommand{\half}{{\textstyle\frac{1}{2}}}
\newcommand{\eqn}[1]{(\ref{#1})}
\newcommand{\npb}[3]{ {\bf Nucl. Phys. B}{#1} ({#2}) {#3}}
\newcommand{\pr}[3]{ {\bf Phys. Rep. }{#1} ({#2}) {#3}}
\newcommand{\prl}[3]{ {\bf Phys. Rev. Lett. }{#1} ({#2}) {#3}}
\newcommand{\plb}[3]{ {\bf Phys. Lett. B}{#1} ({#2}) {#3}}
\newcommand{\prd}[3]{ {\bf Phys. Rev. D}{#1} ({#2}) {#3}}
\newcommand{\hepth}[1]{ [{\bf hep-th}/{#1}]}
\newcommand{\grqc}[1]{ [{\bf gr-qc}/{#1}]}
 
\def\a{\alpha}
\def\b{\beta}
\def\g{\gamma}\def\G{\Gamma}
\def\d{\delta}\def\D{\Delta}
\def\ep{\epsilon}
\def\et{\eta}
\def\z{\zeta}
\def\t{\theta}\def\T{\Theta}
\def\l{\lambda}\def\L{\Lambda}
\def\m{\mu}
\def\f{\phi}\def\F{\Phi}
\def\n{\nu}
\def\p{\psi}\def\P{\Psi}
\def\r{\rho}
\def\s{\sigma}\def\S{\Sigma}
\def\ta{\tau}
\def\x{\chi}
\def\o{\omega}\def\O{\Omega}
\def\k{\kappa}
\def\pa {\partial}
\def\ov{\over}
\def\br{\nonumber\\}
\def\ud{\underline}
\begin{flushright}
SINP/TNP/00-12\\
UUPHY/00/01\\
hep-th/0005051\\
\end{flushright}
\bigskip\bigskip
\begin{center}
{\large\bf 
Supersymmetric configurations in Euclidean Freedman-Schwarz 
model}
\vskip .9 cm
{\sc Swapna Mahapatra$^a$\footnote{e-mail: swapna@iopb.res.in} 
and
Harvendra Singh$^b$\footnote{e-mail: hsingh@tnp.saha.ernet.in}}
\vskip1cm
$^a$ Physics Department, Utkal University\\
Bhubaneswar-751004, India
 \vskip 0.5cm 
$^b$ Theory Division, Saha Institute of Nuclear Physics,\\
1/AF Bidhannagar, Calcutta-700 064, India 

\end{center}
\bigskip
\centerline{\bf ABSTRACT}
\bigskip

\begin{quote}
We study Euclidean $D=4$, $N=4$ gauged $SU(2) \times SU(1,1)$ 
supergravity theory which has been obtained from dimensional 
reduction of $N=1$, $D=10$ supergravity on $S^3 \times AdS_3$. 
We obtain supersymmetric configurations like domain wall, 
electro-vac type of solutions with geometries $E^2 \times 
S^2$, $E^2 \times AdS_2$ and axio-vac type 
$E^1 \times S^3$ solution in this Euclidean 
Freedman-Schwarz (EFS) model. We also show that the Euclidean 
gravitational instantons with nontrivial (anti)self-dual 
$U(1)$ gauge fields are stable vacua preserving one fourth 
of the original supersymmetry.

\end{quote}

\newpage
\section{Introduction}

Recently there has been renewed interest in the study of 
supergravity theories and in particular gauged supergravities
due to the AdS/CFT correspondence \cite{maldacena}
as well as the domain-wall/QFT
 correspondence \cite{boonstra}.  
Extended supergravity theories with $N$ extended local 
supersymmeties have a global $SO(N)$ invariance. 
Gauged supergravity theories
arise when a subgroup of the $R$-symmetry group or the 
automorphism group of the supersymmetry algebra is gauged by 
the vector fields in the graviton supermultiplet. 
There are also other ways to obtain gauged models, like
gauging the isometries of the vector 
multiplet moduli space as well as the hypermultiplet moduli 
space. The procedure of gauging does not change the particle
content of the theory, but it introduces new terms proportional
to the square of the gauge coupling constant in the action. 
So gauging necessarily induces either 
a cosmological 
constant (for $N\leq 3$) in which case, the supersymmetric 
ground state is an 
AdS space or in the presence of scalar fields (for $N\geq 4$), 
a scalar potential which is unbounded from below and it  
may or may not have critical points. Though the potential
is inverted and unbounded from below, in the presence of 
gravity, this becomes a perfectly consistent theory 
having stable
ground state configurations. If a background has killing 
spinors, then one can show that it is a stable background 
by applying Witten's positivity of energy argument 
\cite{witten-gibbons}. Therefore, it is important to 
understand the nature of the ground states of these 
theories as well 
as the relationship between gauged supergravities and consistent 
compactifications/truncations of higher dimensional supergravity 
theories. 

In four dimensions, there are two versions of $N=4$ 
supergravity
theories, one with a global $SO(4)$ symmetry \cite{das}
and the other one
with a global $SU(4)$ symmetry \cite{cremmer}. The equations 
of motion as 
derived from the two versions are equivalent by using field 
redefinition and duality 
transformations. However, when one considers the gauged models 
corresponding to the respective local internal symmetries, one finds
that the two versions are inequivalent. The $N=4$, $SO(4)$ gauged 
model \cite{fischler} has one coupling constant and the 
scalar potential which 
is generated by gauging, is unbounded from below. On the 
otherhand, one can consider the Freedman-Schwarz model 
(FS) \cite{fs}, where
one considers gauging a $SU(2) \times SU(2)$ 
subgroup of $SU(4)$ internal symmetry with two independent
gauge coupling constants. 
Here the scalar potential is again unbounded from below 
and has no critical points. But in both the  
cases, there exist stable vacuum configurations preserving some
amount of supersymmetry. The electro-vac solutions 
\cite{freedman} in gauged 
$SU(2) \times SU(2)$ supergravity is one such well known example.
Other backgrounds in the FS model, like domain walls, strings, 
pure axionic gravity etc have also been recently obtained 
and they preserve either half or one fourth of the 
supersymmetry \cite{singh}. Nonabelian solitons \cite{chams} and
black holes \cite{klemm} as stable vacuum configurations
were also shown to exist. In related work on 
strings in curved backgrounds \cite{anto1}, exact supersymmetric 
solutions of $D=4$ gauged supergravities have been constructed 
by using the techniques of conformal field theory and the 
connection between gauged supergravities and non-critical 
strings have been discussed. 

In this work, we shall concentrate on the Euclidean 
Freedman Schwarz (EFS) model in $D=4$ 
which has recently been obtained by Volkov \cite{volkov}. 
The two theories (FS and EFS)
are different as they are obtained from compactification 
of the ten dimensional theory on different group manifolds
and it is to be noted that they are not just related by 
analytic continuations.   
$D=4$, $N=4$ gauged $SU(2) \times SU(2)$ FS model 
can be embedded 
into \break 
$N=1$ supergravity in ten dimensions as an $S^3 
\times S^3$ 
compactification with the group \break 
manifold being $SU(2) \times 
SU(2)$ \cite{chamseddine}. Previously also, a Kaluza-Klein
(KK) interpretation for the $SU(2) \times SU(2)$ gauged 
supergravity was given in \cite{anto2}, where the model 
was identified as part of the effective $D=4$ field theory 
for the heterotic string theory on $S^3 \times S^3$. These 
two KK interpretations are essentially the same upto consistent
truncations. One can also consider another reduction of the 
$N=1$ ten dimensional theory on the group manifold 
$SU(2) \times
SU(1,1)$ so that the geometry of the internal space-time is 
$S^3 \times AdS_3$ with the signature $(+, +, +, +, +, -)$
and the corresponding four dimensional theory becomes an
Euclidean theory. 
As the scalar curvature of $S^3$ is positive 
and that of $AdS_3$ is negative, the dilaton or equivalently
the scalar 
potential in the corresponding four dimensional theory becomes 
proportional to $g_1^2 - g_2^2$, where, $g_1$ and $g_2$ are the 
gauge coupling constants corresponding to $SU(2)$ and $SU(1,1)$ 
respectively. Since the potential is proportional to 
the square of the difference of the gauge couplings, one 
can consider a variety of cases, where the potential can 
be positive, negative or zero. 
The dimensional reduction on the above 
group manifold is consistent in the sense that for a 
given four dimensional 
configuration which satisfies the four dimensional equations 
of motion and supersymmety variations, the corresponding uplifted
version also satisfies the ten dimensional equations of motion
as well as the supersymmetry variations.

The paper is organized as follows: 
In section 2, we discuss the four dimensional gauged EFS model 
as obtained from the dimensional reduction of the corresponding 
ten dimensional theory. In section 3 and 4, we explicitly 
obtain the new background solutions and illustrate that they 
preserve either half or one fourth of the original supersymmety. 
The Euclidean solutions we have obtained, include the interesting 
cases of domain wall, $E^2 \times S^2$, $E^2 \times AdS_2$, 
$E^1 \times S^3$ where $E^1$ and $E^2$ denote one and two 
dimensional Euclidean spaces. We also show that the four 
dimensional
gravitational instanton solutions \cite{eguchi} like 
Eguchi-Hanson  
is a solution of the EFS model with nontrivial   
(anti)self-dual abelian gauge fields belonging to 
the $U(1)$ of $SU(2)$ and the 
noncompact $SU(1,1)$ groups. In section 5, we summarize our 
results.

\section{The Euclidean Freedman-Schwarz model}

In this section we set our notations and briefly review some 
necessary aspects of the EFS model \cite{volkov} which will 
be necessary for our analysis.  
The field content of the EFS model 
is same as that of the FS model. In the EFS model, 
the four dimensional 
gravity multiplet contains the graviton $E_{\mu}^m$, four 
majorana spin $\frac{3}{2}$ gravitinos $\Psi_{\mu}^I (I = 
1, \ldots 4)$, three nonabelian vector fields $A_{\mu}^a 
(a = 1, 2, 3)$ belonging to $SU(2)$ with gauge coupling $g_1$, 
three nonabelian pseudovector gauge fields $\dot A_{\mu}^a$ 
belonging to 
$SU(1,1)$ group with gauge coupling constant $g_2$, four majorana
spin $\frac{1}{2}$ fields $\chi^I$, the axion ${\bf a}$ and the 
dilaton $\Phi$. Here the Greek indices $\mu, \nu, \ldots$ 
refer to the base space indices and latin indices $m, n, \ldots$
refer to the tangent space indices. The bosonic part of the ten 
dimensional theory contains the metric $\hat g_{MN} (M, N, \ldots 
= 1, ... 10)$, the three form antisymmetric tensor $\hat H_{MNP}$
and the dilaton $\hat\Phi$. The fermionic field contents are 
the ten dimensional gravitino $\hat \Psi_M$ and the gaugino 
$\hat \chi$. We consider vanishing spinor background fields, 
however their supersymmetric variations do not vanish and
they are important for our considerations. 

The bosonic part of the ten dimensional action corresponding to 
$N=1$ supergravity is given by, 

\bea 
S_{10} & = & \int {\sqrt {- \hat g}}~~ d^{10}\hat x 
\bigg (\frac{1}{4}
\hat R - \frac{1}{2} \partial_M \hat\Phi \partial^M \hat\Phi
- \frac{1}{12} e^{- 2 \hat\Phi} {\hat H}_{MNP} 
{\hat H}^{MNP} \bigg )
\eea
where $\hat R$ is the curvature scalar in $D=10$. The equations 
of motion following from this action are given by, 

\bea
\hat\nabla_M \hat\nabla^M \hat\Phi + \frac{1}{6} e^{-2 \hat\Phi}
{\hat H_{MNP}}{\hat H^{MNP}} & = & 0
\\
\hat\nabla_M (e^{-2\hat\Phi} {\hat H}^{MNP}) & = & 0
\\
{\hat R}_{MN} - 2 \pa_M \hat\Phi \pa_N \hat\Phi - e^{-2\hat\Phi}
{\hat H}_{MPQ} {\hat H}_N^{PQ} + \frac{1}{12} e^{-2\hat\Phi}
{\hat g}_{MN} {\hat H}_{PQS} {\hat H}^{PQS} & = & 0
\eea
 
The dimensional reduction of the above ten dimensional theory 
in terms of suitable parametization has been discussed in a 
recent paper by Volkov \cite{volkov}.  
The corresponding four dimensional equations 
of motion 
for metric, dilaton, axion and gauge fields are respectively
given by, 
\bea
&& R_{\m\n} - 2 \pa_{\m}\Phi \pa_{\n}\Phi + 2 e^{-4\Phi}\pa_{\m}
{\bf a}\pa_{\n}{\bf a} \nnu\\
&&\hskip.5cm -2 e^{2\Phi}\left [ \eta^{(1)}_{ab}
\left (F^a_{\m\l} {F}_{\n}^{b \l} - \frac{1}{4} g_{\m\n} 
F^a_{\l\rho} F^{b\l\rho} \right ) + \eta^{(2)}_{ab} 
\left ({\dot F^a}_{\m\l} {\dot F}_{\n}^{b \l} 
- \frac{1}{4} 
g_{\m\n} {\dot F^a}_{\l\rho} {\dot F}^{b\l\rho} \right ) 
\right ] \nnu\\ 
&&\hskip2cm - 2 g_{\m\n} U(\Phi) =  0
\\
&&\nabla_{\m}\nabla^{\m}\Phi - 2 e^{-4\Phi}\pa_{\m}{\bf a}
\pa^{\m}{\bf a} - \frac{1}{2} e^{2\Phi}\left [ 
 \eta^{(1)}_{ab} F^a_{\m\n} F^{b\m\n} + 
\eta^{(2)}_{ab} \dot F^a_{\m\n} \dot F^{
b\m\n} \right ] + 2 U(\Phi) = 0
\\
&&\nabla_{\m} (e^{-4\Phi}\nabla^\m {\bf a}) + \frac{1}{2}
\left [ \eta^{(1)}_{ab} \star F^a_{\m\n}
F^{b\m\n} +  \eta^{(2)}_{ab} \star \dot F^a
_{\m\n} \dot F^{b\m\n} \right ]  =  0
\\
&&\nabla_{\rho} (e^{2\Phi} F^{a\rho\m}) + g_1e^{2\Phi}{f^a}_{bc}
{A^b}_{\rho} F^{c\rho\mu} - 2 \star F^{a\m\rho} \pa_{\rho}
{\bf a} = 0
\\
&&\nabla_{\rho} (e^{2\Phi} \dot F^{a\rho\m}) +g_2 e^{2\Phi}
{{\dot f}^a}_{bc} {{\dot A}^b}_{\rho} \dot F^{c\rho\m} - 
2 \star \dot F^{a\m\rho} \pa_{\rho}{\bf a} = 0
\eea
where $U(\Phi)$ is the dilaton potential given by, 
\bea 
U(\Phi) & = & - \frac{1}{8} (g_1^2 - g_2^2) e^{-2\Phi}
\eea
The structure constants are given by, 
\be
{f^c}_{ab} = \eta^{(1)cd} \ep_{dab} ~~; \qquad \qquad 
{{\dot f}^c}_{ab} = \eta^{(2)cd} \ep_{dab}
\\
\ee
$\ep_{abc}$ is the antisymmetric tensor, $\eta^{(1)}_{ab}$
and $ -\eta^{(2)}_{ab}$ are the cartan metrics corresponding 
to $SU(2)$ and $SU(1,1)$ respectively, where 
$\eta^{(1)}_{ab} = diag(1, 1, 1)$ and $\eta^{(2)}_{ab}
= diag(1, 1, -1)$. The dual field strengths in the four 
dimensional Euclidean theory are defined as,
\bea
\star F^a_{\m\n} & = & \frac{1}{2} {\sqrt g} \ep_{\m\n\l\rho}
F^{a\l\rho} 
\eea
and,
\bea
\star \dot F_{\m\n}^a & = & \frac{1}{2} {\sqrt g} \ep_{\m\n
\l\rho} \dot F^{a\l\rho}
\eea

The above equations of motion can be obtained from the four 
dimensional Euclidean Freedman-Schwarz action,
\bea
&& S_4  =  \int {\sqrt g}~~ d^4 x \bigg [ \frac{R}{4}
- \frac{1}{2} \pa_{\m}\Phi\pa^{\m}\Phi + \frac{1}{2}
e^{-4\Phi}\pa_{\m}{\bf a}\pa^{\m}{\bf a} - \frac{1}{4}
e^{2\Phi} \left (  \eta^{(1)}_{ab} 
F^a_{\m\n} F^{b\m\n} +  \eta^{(2)}_{ab}
\dot F^a_{\m\n} \dot F^{b\m\n} \right ) \nnu\\
&&\hskip1cm - \frac{1}{2} {\bf a} \bigg (
\eta^{(1)}_{ab} \star F^a_{\m\n} F^{b\m\n} + 
 \eta^{(2)}_{ab} \star \dot F^a_{\m\n}
\dot F^{b\m\n} \bigg ) + \frac{1}{8} (g_1^2 - g_2^2)
e^{-2\Phi} \bigg ]
\eea

Similarly the ten dimensional spinors can also be consistently 
reduced and the four dimensional supersymmetry variations are 
given by, 
\bea 
&&\delta \x = \left ( \frac{1}{\sqrt 2} \g^{\m}\pa_{\m}\Phi 
- \frac{1}{\sqrt 2} e^{-2\Phi} \g_5 \g^{\m}\pa_{\m}{\bf a}
\right ) \ep  
+ \frac{1}{2} e^{\Phi} \left ( \frac{1}{2} \eta^{(1)}_
{ab} \g^{\a} \g^{\b} F^a_{\a\b} {\bf \a}^b - \frac{1}{2}
\g_5 \eta^{(2)}_{ab} \g^\a \g^\b \dot F^a_{\a\b} {\bf 
\dot \a}^b
\right ) \ep \nnu\\ 
&&\hskip1cm  + \frac{1}{4} e^{-\Phi} (g_1 - g_2 \g_5) \ep,\nnu\\ 
&&\delta \Psi_{\m} = \bigg ( \pa_{\m} + \frac{1}{4}
\omega^{\a\b}_{\m} \g_{\a}\g_{\b} - \frac{g_1}{2} 
\eta^{(1)}_{ab} {\bf \a}^a A^b_{\m} + \frac{g_2}{2} 
\eta^{(2)}_{ab} {\bf \dot \a}^a \dot A^b_{\m} + 
\frac{1}{2} e^{-2\Phi} 
\g_5\pa_{\m} {\bf a} \bigg )\ep + \nnu\\ 
&&\hskip1cm \frac{1}{2{\sqrt 2}}
e^{\Phi} \left ( \eta^{(1)}_{ab} F^a_{\l\n} {\bf \a}^b + \g_5
\eta^{(2)}_{ab} \dot F^a_{\l\n} {\bf \dot \a}^b \right )
\g^{\l} \g^{\n} \g_{\m} \ep + \frac{1}{4{\sqrt 2}} e^{-\Phi}
(g_1 + g_2 \g_5) \g_{\m} \ep 
\eea
where $\ep$ is the Majorana spinor corresponding to 
the supersymmetry transformation parameter.  
Here, $\g_5 = - \g^0\g^1\g^2\g^3$ and $\{\g_5, \g^{\a}\} = 0$.
$\g^{\a}$ are the four dimensional tangent space gamma matrices
satisfying the usual anticommutation relation
$\{\g^{\a}, \g^{\b}\} = 2 \eta^{\a\b}$ with $\eta^{\a\b} = 
diag(+1, +1, +1, +1)$. $\omega^{\a\b}_{\m}$ are the spin 
connections
and ${\bf \a}^a$ and ${\dot \a}^a$ are the $4 \times 4$ 
matrices which generate the Lie algebra of the group 
$SU(2)$ and $SU(1,1)$ respectively with the properties,

\bea
{\bf \a}^a {\bf \a}^b & = & - \ep^{abc} \eta^{(1)}_{cd} 
{\bf \a}^d - \eta^{(1)ab}
\\
{\bf \dot \a}^a {\bf \dot \a}^b & = & -\ep^{abc} 
\eta^{(2)}_{cd} {\bf \dot \a}^d 
+ \eta^{(2)ab} 
\\
\big [{\bf \a}^a , {\bf \dot \a}^a \big ] & = & 0
\eea
The corresponding matrix representation is given by, 
\be
{\bf \a}^a = i \tau^a \otimes I_2 ~~(a = 1,2,3)~;
\quad {\bf \dot \a}^b = I_2 \otimes \tau^b ~~
(b = 1,2)~; \quad {\bf \dot \a}^3 = i I_2 \otimes \tau^3
\ee
where $\tau$'s are Pauli matrices. 

In next section, we shall show that one can construct
many interesting stable new vacua for the EFS model 
which are supersymmetric and are consistent with the four
dimensional background equations of motion.

\section{Supersymmetric configurations in EFS model}
In the examples below, we choose specific $U(1)$ directions
thereby spontaneously breaking $SU(2)$ to $U(1)$ and 
in the noncompact case, $SU(1,1)$ to $U(1)$, similar to 
Freedman-Gibbons electro-vac solutions with constant 
dilaton. This corresponds 
to setting the other two gauge fields of $SU(2)$ triplets or 
the $SU(1,1)$ triplet to zero vacuum expectation value.
So whenever we have nonzero gauge fields, they are 
basically abelian. 

\subsection{Euclidean domain walls}

First we consider the four dimensional Euclidean domain 
wall obtained by analytically continuing the Lorenztian 
domain walls \cite{singh} with the field configurations,

\bea 
&&d s^2  =  U(y) (d t^2 + d x_1^2 + d x_2^2) + 
U^{-1}(y) d y^2 , \nnu\\ 
&&\Phi  =  \frac{1}{2}\ln U(y), 
\qquad U(y)  =  m|y - y_0| , \nnu\\
&&A^a_{\m} = 0, \qquad \dot A^a_{\m} = 0, \qquad {\bf a} = 0,
\eea
This background is singular at $y = y_0$. Since there is no 
other matter field present here other than dilaton, the 
above configuration 
represents pure dilaton gravity in Euclidean space. 
We now study the supersymmetric properties of this background.  
For $g_1 \neq 0$ (in fact the gauge coupling constant and the 
mass parameter are related by $g_1^2 = 2 m^2$) and $g_2 = 0$, 
if we 
substitute the above background in the supersymmetry equations,
we find that the fermionic variations vanish provided the 
supersymmetry parameters satisfy, 
 
\be
\ep = - \g_3 \ep , ~~\qquad \ep = U(y)^{\frac{1}{4}} \ep_0
\ee
where $\ep_0$ is a constant spinor. These conditions 
break half of the supersymmetry. Thus we see 
that there exist nontrivial killing spinors preserving 
$N=2$ supersymmetry for pure dilatonic Euclidean domain 
wall background.

\subsection{$E^2 \times S^2$} 
Here we consider the case where   
dilaton is constant. We also take one of the $U(1)$ gauge fields 
of the $SU(2)$ part to be nonvanishing and the geometry  
as that of $E^2 \times S^2$. Corresponding field configurations 
are given by, 
\bea 
&&ds^2 = d\psi^2 + d\x^2 + \frac{1}{B} \bigg ( d\theta^2 + 
\sin^2\theta d\phi^2 \bigg ) \nnu\\
&&F^a = \delta^{a3} Q \sin\theta d\theta \wedge d\phi~~; 
\qquad \Phi =
\Phi_0  = constant \nnu \\ 
&& \dot F = 0, \qquad {\bf a} = constant, \qquad g_1 = 0
\eea
where $\frac{1}{\sqrt B}$ is the constant radius of the
two sphere. This configuration satisfies the equations 
of motion with $Q = \frac{1}{g_2}$. The above solution 
is analogous to the magnetic solution in the charged
Nariai black hole background \cite{nariai}, 
but here we are in the 
Euclidean space with $F^2 = 2 B^2 Q^2$ which is 
strictly positive.
   
Next, we discuss the supersymmetric property of this 
background. The only nonvanishing spin connection is 
$\o^{23}_{\phi} = - \cos\theta$. 
The components of the killing spinor equations are given by, 
\bea
&&\pa_{\psi}\ep = 0 \\
&&\pa_{\chi}\ep = 0 \\
&&\pa_{\theta}\ep + \frac{1}{2} \g_5 \g_2 \ep = 0 \\
&&\pa_{\phi}\ep - \frac{1}{2} \cos\theta \g_2\g_3 \ep 
+ \frac{1}{2} \sin\theta \g_5\g_3 \ep = 0
\eea 
with the condition
$$\left ( \g_5 \g_2\g_3 {\bf \a}^3 - 1 \right )\ep=0.$$ 

The complete set of killing spinors which are the solution to the above
equations are
\be
\ep = e^{- \frac{1}{2}\theta\g_5\g_2} e^{-\frac{1}{2}\phi
\g_3\g_2}\left( {\g_5\g_2 \g_3 {\bf \a}^3 + 1\ov2}\right )\ep_0
\ee
where $\ep_0$ is some constant spinor and ${\bf \a}^3$ is 
in $SU(2)$. The operator 
$\left [ \g_5 \g_2\g_3 {\bf \a}^3 + 1 \right ]$ 
acts as a projection operator, hence breaks 
$\frac{1}{2}$ supersymmetry. 
 
\subsection{$E^2 \times AdS_2$}

Here we consider the geometry $E^2 \times AdS_2$ with constant 
dilaton and the nonvanishing $U(1)$ gauge field corresponding
to the noncompact part of $SU(1,1)$. This choice of gauge field 
is necessary  so that the background equations of motion 
are satisfied. 

The field configurations are given by, 
\bea 
&&d s^2 = d\psi^2 + d\chi^2 + \frac{1}{B}\bigg (r^2 d t^2 + 
\frac{dr^2}{r^2}\bigg ) \nnu\\
&&\dot F^a = \delta^{\dot a3} Q dt \wedge dr ~~ , \qquad 
\Phi = \Phi_0 = 
constant \nnu \\
&&F = 0, \qquad {\bf a} = constant, \qquad g_2 = 0 
\eea
where $\frac{1}{\sqrt B}$ corresponds to the radius of the 
AdS space. The above background fields satisfy the equations
of motion with $Q = \frac{1}{g_1}$. The only nonzero spin 
connection is given by $\o^{23}_t = r$. 
Considering that the fermionic supersymmetry variations vanish, 
one gets the following
equations for the $\psi$, $\x$, $t$ and $r$ components for the 
killing spinor: 
\bea
&&\pa_{\psi}\ep = 0 \\
&&\pa_{\x}\ep = 0  \\
&&\pa_t \ep + \frac{r}{2} \g_2 \g_3 \ep + \frac{1}{2}
r \g_2 \ep = 0  \\
&&\pa_r \ep + \frac{1}{2r} \g_3 \ep = 0 
\eea
The projector is given by $\left [ \g_5\g_2\g_3{\bf \dot \a}^3 
+ 1 \right ]$ where ${\bf \dot \a^3}$ is along the noncompact 
direction in $SU(1,1)$. The killing spinors are
\bea 
&&\ep = {1\ov2}r^{\frac{1}{2}} \left (\g_5 \g_2 \g_3 
{\bf \dot \a}^3 + 
1 \right ) \ep_{-} + {1\ov2}\left [ r^{-\frac{1}{2}} - 
r^{\frac{1}{2}}
\g_2 t \right ] \left ( \g_5 \g_2 \g_3 {\bf \dot \a}^3 + 1 
\right ) \ep _{+}
\eea
where $\ep_{\mp}$ are constant spinors and they satisfy the 
conditions $\g_3 \ep_{\mp} = \mp \ep_{\mp}$. 
So as before, this background preserves one half of the 
supersymmetry. 

These last two EFS solutions are analogous to 
the electro-vac solution in FS model. The later one, 
$E^2\times AdS_2$, can
be mapped to the Lorentzian section by applying the 
transformations as
dicussed in \cite{volkov}.

\subsection{$E^1 \times S^3$}

Here, we would like to obtain a background analogous to 
the pure axionic
gravity solution  in FS model \cite{singh}. So, we take 
nontrivial axion field while the dilaton as well as  
gauge fields are vanishing.
The field configuration is given by,
\bea 
&&ds^2 = d \psi^2 + \frac{1}{Q} \bigg ( d\x^2 + 
\sin^2\x \bigg ( d\theta^2 + \sin^2\theta d\phi^2 \bigg )
\bigg ) \nnu \\
&& {\bf a} =\pm {\sqrt Q}~\psi ; \qquad \Phi = 0 \nnu \\
&& A = 0; \qquad \dot A = 0; \qquad g_1 = 0
\eea
The equations of motion of these background fields are 
consistent with $Q = \frac{g_2^2}{8}$. To study the 
supersymmetry 
properties, we need to calculate the spin connections 
on three sphere. 
The nonzero components are given by, 
\bea 
&&\o^{21}_{\theta} = \cos\chi ; \qquad \o^{31}_{\phi} = 
\cos\chi\sin\theta ; \qquad \o^{32}_{\phi} = \cos\theta
\eea
The projector is given by, 
\be
\ep = - \g_0 \ep,
\ee 
and the components of the killing equations are given by, 
\bea 
&&\pa_{\psi}\ep = 0 \\
&&\pa_{\chi}\ep + \frac{1}{2} \g_5 \g_1 \ep = 0 \\
&&\pa_{\theta}\ep + \frac{1}{2} \cos\chi \g_2 \g_1 \ep 
+ \frac{1}{2}\sin\chi \g_5 \g_2 \ep = 0  \\
&&\pa_{\phi}\ep + \frac{1}{2} \cos\chi\sin\theta \g_3\g_1\ep 
+ \frac{1}{2} \sin\chi\sin\theta\g_5\g_3 \ep + 
\frac{1}{2} \cos\theta\g_3\g_2 \ep = 0
\eea
The complete solution of the killing spinor equation
is given by, 
\bea
&& \ep = e^{-\frac{1}{2}\chi\g_5\g_1} e^{-\frac{1}{2}
\theta \g_2 \g_1} e^{-\frac{1}{2}\phi \g_3 \g_2} 
\bigg [{ \g_0 + 1 \ov 2}\bigg ] \ep_0
\eea
Hence this choice of field configurations breaks $\frac{1}{2}$ 
of the supersymmetry. 

We find that $E^1 \times AdS_3$ background can also be a 
solution 
but then one has to consider imaginary axion field to 
solve of the background equations of motion.  

\section{Gravitational Instantons}

In this section we consider gravitational instantons 
which are solutions in Euclidean gravity. It has been noted 
in \cite{volkov} that with vanishing dilaton, 
axion, gauge fields and for $g_1=g_2$,
the flat gravitational instantons 
(cosmological constant being zero) are vacua of EFS model. 
Here we show that even in the presence of (anti)self-dual 
gauge fields, the 
Eguchi-Hanson instanton \cite{eguchi} satisfying the flat 
space Einstein equations is a consistent background of this 
EFS model and it preserves certain 
fraction 
of the supersymmetry.  

This is one of the examples, where we keep both the gauge 
coupling constants and we take them to be equal, $g_1 = g_2$. 
We take nonzero $U(1)$ gauge field belonging to the 
noncompact 
part of $SU(1,1)$ and the nonvanishing  $U(1)$  gauge 
field of the $SU(2)$ part could be any of the triplet (let us 
choose $A^3$ to be nonzero). 
The field configuration is given by, 
\bea
&&ds^2 = \frac{dr^2}{1 - \frac{a^4}{r^4}} + \frac{r^2}{4}
\bigg (d\theta^2 + \sin^2\theta d\phi^2 \bigg ) + 
\frac{r^2}{4} \bigg (1 - \frac{a^4}{r^4} \bigg ) 
{\bigg ( d\psi + \cos\theta d\phi \bigg )}^2 \nnu \\
&& F = \frac{2}{r^4} \bigg (e^3 \wedge e^0 + 
e^1 \wedge e^2 \bigg ) = \dot F \nnu \\
&& \Phi = 0; \qquad {\bf a} = 0; \qquad g_1 = g_2
\ll{42}
\eea
where the vierbeins are  
\bea
&& e^0 = \frac{dr}{\sqrt{1 - \frac{a^4}{r^4}}} ; \qquad 
\qquad \qquad e^1  = \frac{r}{2} \bigg (\sin\psi d\theta - \sin
\theta\cos\psi d\phi \bigg ) \nnu \\
&&e^2 = \frac{r}{2}\bigg ( -\cos\psi d\theta - \sin\theta
\sin\psi d\phi \bigg ); \quad e^3 = \frac{r}{2}
{\sqrt{1 - \frac{a^4}{r^4}}} \bigg ( d\psi + \cos\theta d\phi
\bigg ).
\ll{43}
\eea
One can immediately note that the gauge field strengths 
in \eqn{42} are
anti-self-dual. 
The spin connections which are also anti-self-dual can be 
calculated from
\eqn{43} and these are 

\bea
&&\o^{10}_{\theta}=\o^{23}_\theta = \frac{1}{2} {\sqrt{1 -
\frac{a^4}{r^4}}} 
\sin\psi~~; \quad ~~~\o^{10}_{\phi} =\o^{23}_{\phi} = 
-\frac{1}{2}
{\sqrt{1 -
\frac{a^4}{r^4}}} \sin\theta\cos\psi \nnu \\
&&\o^{20}_{\theta} =\o^{31}_{\theta} = -\frac{1}{2} {\sqrt{1 -
\frac{a^4}{r^4}}}
\cos\psi~~; \quad \o^{20}_{\phi} =\o^{31}_{\phi} = 
-\frac{1}{2} {\sqrt{1 - 
\frac{a^4}{r^4}}} \sin\theta\sin\psi \nnu \\
&&\o^{30}_{\psi} =\o^{12}_{\psi} = \frac{1}{2}\big ( 1 + 
\frac{a^4}{r^4} 
\big )~~; \qquad \qquad ~\o^{30}_{\phi} = \o^{12}_{\phi} =
\cos\theta 
\o^{30}_{\psi}.
\eea  
With the above choice of background fields, 
the supersymmetry variations give the projector conditions, 
\bea
&& \bigg (1 - \g_5 \bigg )\ep = 0 \nnu \\
&& \bigg ( {\bf \a}^3 + {\bf \dot\a}^3 \bigg ) \ep = 0
\ll{123} 
\eea
With these projectors, the killing spinor equations are 
really simplified: 
\bea
&&\pa_\m~\ep = 0. 
\ll{124}
\eea
So the killing spinors are independent of $ r, \theta, 
\phi, \psi$. Because of the twin supersymmetric 
conditions, the solution preserves $\frac{1}{4}$ of the 
supersymmetry. However, once the gauge field backgrounds are
switched off the second condition in eq.\eqn{123} will drop out
and the pure gravitational instanton background will become half
supersymmetric.

\section{Summary}

In this work, we have obtained stable vacuum configurations 
in $D=4$, $N=4$ $SU(2) \times SU(1,1)$ gauged supergravity 
theory (EFS model) which has been obtained from dimensional 
reduction of $D=10$, $N=1$ supergravity on $S^3 \times 
AdS_3$. We have obtained stable domain wall solutions 
preserving half of the supersymmetry. We have then considered 
vacua like $E^2 \times S^2$, $E^2 \times AdS_2$
with constant dilaton and nontrivial $U(1)$ gauge 
fields which are analogous  
to the electrovac solutions in FS model
preserving half of the supersymmetry. We also obtained 
geometry like $E^1 \times S^3$ with nontrivial axion and 
vanishing dilaton field preserving half of the supersymmetry. 
Finally we obtained  Euclidean 
gravitational instanton, namely Eguchi-Hanson  
instanton with nontrivial abelian gauge fields and 
vanishing dilaton and axion 
fields as a consistent vaccum configuration breaking
one fourth of the supersymmetry. The dilaton potential 
in this case vanishes as the two gauge coupling constants 
are equal. So our findings of a rich variety of vacua for 
the EFS model makes the theory more interesting and worth 
exploring further.

%

\end{document}